\documentclass[preprint2]{aastex}

\def\deg{\hbox{$^\circ$}}
\def\la{\mathrel{\hbox{\rlap{\hbox{\lower4pt\hbox{$\sim$}}}\hbox{$<$}}}}
\def\ga{\mathrel{\hbox{\rlap{\hbox{\lower4pt\hbox{$\sim$}}}\hbox{$>$}}}}

\begin{document}
\title{Physics Results From Alpha Magnetic Spectrometer \\1998 Shuttle Flight}
\author{{\em Proceeding of the 7-th Taiwan Astrophysics Workshop, \\
National Central University, Taiwan, \\ Oct. 18,2000} \\ \vspace{0.5cm}
M. A. Huang}
\affil{Institute of Physics, Academia Sinica, Taipei, 11529, 
	Taiwan, R.O.C.}
\email{huangmh@phys.sinica.edu.tw}

\begin{abstract}
The Alpha Magnetic Spectrometer (AMS) is a particle detector designed to 
detect antimatter. During the 10-day test flight on the space shuttle in 
June 1998, AMS detected $10^8$ events. Upon analysis, no antimatter was 
found and the antimatter limit was reduced to $1.1\times10^{-6}$. 
The proton spectrum shows some differences with the cosmic ray flux used in 
atmospheric neutrino simulation. A large amount of protons, positrons, and 
electrons were found below the geomagnetic rigidity cutoff. The energy of
 these particles are as high as several GeV, one order of magnitude higher 
than any previously measured energy in radiation belts. These particles also 
exhibit many interesting features. This paper reviews the results in the four
 published papers of the AMS collaboration and provides explanation for some
features of the albedo particles. \\
\end{abstract}

\section{Introduction}
The Alpha Magnetic Spectrometer is a space borne charged particle detector 
(Ahlen 1994) and will be installed on the International Space 
Station in 2003 for three years. 
Its aims are to detect the antimatter and dark matter, and to perform 
precision measurement of primary cosmic rays. In June 1998, the detector 
performed an engineering test run on board the space shuttle mission STS-91 for
 10 days and recorded approximately $10^8$ events. The physics results had been
 analyzed and published. This review begins with a short introduction to the 
AMS. Section 2 describes the physics of charged particles in the geomagnetic 
field. The physics results from the AMS 1998 shuttle flight are presented in 
the rest of this paper. The final section summarizes all the results.

\subsection{AMS physics goals}
From our current understanding of elementary particle physics, energy 
always materializes to equal amount of matter and antimatter. If antimatters 
exist in the universe, then their existence could be detected by either 
indirect search using gamma ray line spectrum, which comes from the 
annihilation of antimatter and matter; or by the direct search using cosmic 
rays detector. On the other hand, if antimatters are absent in the universe, 
then CP-violation and baryon nonconservation must be observed. However, all 
such searches have been negative (references 1-7 of Alcaraz 1999). There are no
 positive evidences supporting the existence or absence of antimatter. A 
direct detection of anti-nuclei such as anti-helium or anti-carbon could 
signal the existence of antimatter. The AMS is an accurate, large acceptance 
magnetic spectrometer, which will be installed on the International Space 
Station for three years. The large acceptance and long duration will reduce the
 anti-helium limit to $10^{-9}$.

Dark matter is another unsolved puzzle of the universe. One of the candidates 
of dark matter, WIMP (Weakly Interacting Massive Particle), could annihilate 
in the halo of galaxy and produce an excess of positrons (Hardung \& Ramaty 
1987; Aharonian \& Atoyan 1991; Dogiel \& Sharov 1990; Tylka 1989; 
Turner \& Wilczek 1990; Kamionkowski \& Turner 1991). The AMS can make 
indirect search of WIMP through the detection of positrons.

The majority of particles detected by the AMS are cosmic rays. The large 
acceptance and multiple sub-detectors of AMS can make precise measurements of 
cosmic rays flux and composition. These data are useful for studies on cosmic 
rays and atmospheric neutrino simulation (Honda et al. 1995, Gaisser 1999).

\subsection{AMS01 space shuttle flight}
The AMS collaboration consists of 32 institutions from 13 
nations\footnote{Three institutions from Taiwan 
participated in this project. The Chung Shan Institute of Science and 
Technology designed and manufactured the electronics boards, while the 
Academia Sinica and National Central University were involved in data 
acquisition system and data analysis.}.
In June 1998, a prototype detector, called AMS01, was flown in 
space shuttle Discovery on flight STS-91. During this 10-day test flight, 
AMS01 gathered approximately $10^8$ events. The various components of the AMS 
are listed in Table \ref{tab:ams01} and shown in Figure 1. The 
notation of the coordinate system is as follows. $\widehat{x}$ is the magnetic 
field direction, non-bending direction; $\widehat{y}$ is the bending direction,
 and $\widehat{z}$ is the cylindrical axis.

\begin{figure}[htb]
\plotone{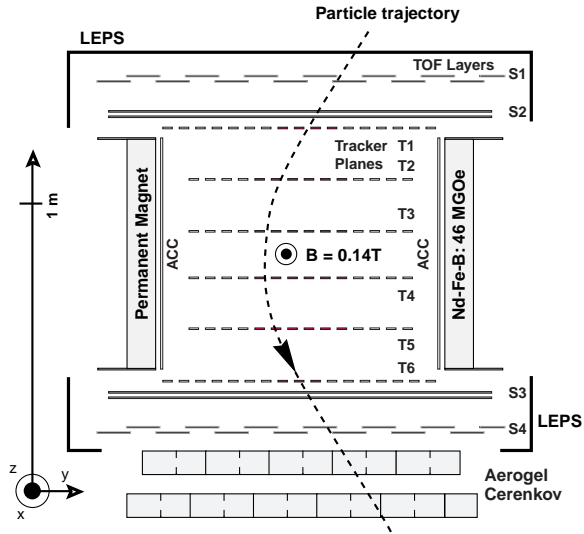}
\label{fig:ams_det}
\caption{Schematic view of the AMS01 detector.}
\end{figure}

\begin{deluxetable}{cl}
\tablecaption{The components of the AMS01 detector and their properties. \label{tab:ams01}}
\tabletypesize{\scriptsize}
\tablewidth{0pt}
\tablehead{\colhead{Components} & \colhead{Descriptions}}
\startdata
	& 1.9 tons of Nd-Fe-B permanent magnet, \\
Magnet	&	Inner radius 1115 mm, length 800mm, \\
	&	Dipole field perpendicular to cylindrical axis, \\
	&	Analyzing power $\mbox{BL}^2=0.14 Tm^2$. \\
\tableline
 	& 6 planes double-side silicone tracker, \\
Tracker	&	resolution in bending direction $\sigma_y=20\mu m$, \\
	&	non-bending direction $\sigma_x=33\mu m$, \\
	&	Measure rigidity and charge. \\
\tableline
 	& 4 planes plastic scintillators, two above magnet, two under magnet.\\
Time of Flight	& Resolution 120psec \\
(TOF)	&	Measure velocity and charge \\
\tableline
Aerogel Cherenkov  & Refractive index 1.035 \\
counter	&	Separation of leptons/hadrons at energy $< 1$ GeV \\
\tableline
Anticoincidence	& Plastic scintillators surrounding the inner surface of magnet.\\
counter & Veto the events penetrating into magnet or interacting in the magnet. \\
\tableline
Low energy  & Carbon fiber 6 mm thick, density $1.3 \mbox{gm/cm}^3$ \\
particle shield	&	Shield from low energy particles \\
\tableline 
 	& Inclination angle $51.7\deg$ \\
AMS01 flight &	Altitude 350 to 390km \\
	&	Data taking time: 184 hours \\
 \enddata
\end{deluxetable}

\section{Introduction to physics of charged particles in the geomagnetic field}
 Since cosmic rays are charged particles, their trajectories are bent by the 
magnetic field. The geomagnetic field affects the arrival directions of cosmic 
rays, blocks lower energy cosmic rays, and traps some low energy particles in 
radiation belts.

\subsection{Geomagnetic field}
The geomagnetic field consists of two parts, the internal field (main field) 
produced by magnetic moments of the Earth and the external field driven by the 
solar wind plasma. The internal and external fields merge at distance 
$>10R_E$, where $R_E$ is the mean earth radius 6371.2 km, forming the 
magnetosphere. Although the AMS flight was at low altitude of 380km, the 
magnetic field lines passing through the AMS path can reach an altitude of 
approximately 6.4 $R_E$, well inside the magnetosphere. Therefore, only the 
internal field is considered in the AMS physics analysis. 

Although the main field can be simplified as a dipole field, the error 
could be as big as $30\%$ in some regions. In realistic field, a series 
of spherical harmonics were used to fit to the measured value.
\begin{eqnarray*}
  V&=&R_E \sum^{\infty}_{n=0}(\frac{R_E}{r})^{n+1} \sum_{m=0}^{n}P_n^m(\cos\theta) \times\\
& &   (g_n^m \cos{m\phi} + h_n^m \sin{m\phi}) 
\end{eqnarray*}
Where the $P_n^m(\cos\theta)$ is the associated Legendre function with Schmidt 
normalization of the degree n and order m. $g_n^m$ and $h_n^m$ are 
Gaussian coefficients determined by the magnetic field. The coefficient $n=1$ 
corresponds to the dipole term.

Every five years since 1945, the International Geophysics Union publishes the 
fitted coefficients to the 10th degree. This field model is called the 
International Geomagnetic Reference Field (IGRF). Also, the US Department of 
Defense fits the coefficients to the 12th degree, this field model is called 
the World Magnetic Model (WMM). Those two models are slightly different 
(Huang 2001) and their details can be found on the 
web\footnote{http://nssdc.gsfc.nasa.gov/space/model/ and \\
http://www.ngdc.noaa.gov/seg/potfld/magmodel.shtml}.

Along the west coast of South America and southern Atlantic Ocean, the magnetic
 field is the weakest on the Earth surface. Because of high radiations, many 
satellites experience some troubles when flying over these regions. The 
phenomenon in this region is known as the {\em "South Atlantic Anomaly"} (SAA). 
Because of high deadtime when the AMS flies through the SAA, the data taken in 
the SAA are excluded in the data analysis. 

For magnetic coordinates, the dipole coordinates were relative to the dipole 
axis. In realistic field, Corrected GeoMagnetic coordinates (CGM,  Gustafsson 
1992) were commonly used by space scientists and geophysicists. The $0\deg$ of 
CGM longitude passes through the SAA area, north and south magnetic poles. 
Figure 2 shows the contour line of magnetic field strength and the 
CGM coordinates at the Earth surface. The notation of magnetic latitude used 
in AMS publications is $\theta_m$. To avoid confusion with the zenith angle 
and to be consistent with common practice, $\lambda_m$ is used in this 
paper. The symbol of magnetic longitude is $\phi_m$.

\begin{figure}[htb]
\plotone{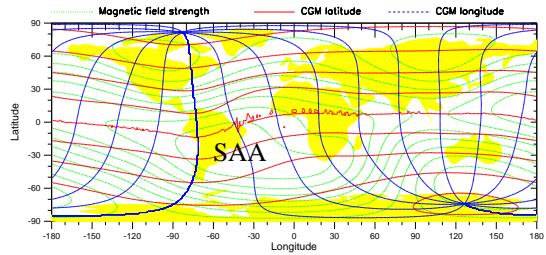} 
\label{fig:cgm}
\caption{The magnetic field and coordinates used in AMS01. The 
dotted contour lines are the magnetic field strength at the Earth surface, and 
the solid/dash lines are the magnetic latitude/longitude. The South Atlantic 
Anomaly region is marked in this figure.}
\end{figure}

\subsection{Rigidity cutoff}
The geomagnetic field shields the Earth from cosmic rays bombardment.
 Below a threshold, the rigidity cutoff, the cosmic rays cannot penetrate 
through the geomagnetic field into the lower atmosphere. Also particles with 
rigidity below the threshold cannot escape from the geomagnetic field, i.e. 
particles are trapped by the geomagnetic field. In dipole field, the rigidity 
cutoff can be expressed in analytical form, called Str\"{o}mer cutoff 
(Str\"{o}mer 1930):
\[
R_{cutoff}=\frac{M\cos^4{\lambda}}{r^2(1+\sqrt{1-\cos{\alpha}
                                         \cos^3{\lambda}})^2} 
\]
\begin{equation}
R_{cutoff} = \frac{59.6\ \cos^4{\lambda}\ (GV/c)}{(r/R_E)^2(1+
	   \sqrt{1-\cos{\gamma}\cos^3{\lambda}})^2} 
\end{equation}
\begin{center}
\begin{tabular}{rl}
 $M$: & Dipole moment; \\ 
$\lambda$: & Magnetic latitude; \\
 $r$: & Radial distance; \\
 $\gamma$: & Incident angle from the west. \\
\end{tabular}
\end{center}

In realistic field, the cutoff cannot be formulated as a simple formula. The 
cutoff becomes a band of intermittent transition, which is called penumbra. 
Numerical calculation is needed to study the cutoff in realistic field. Figure 
3 shows the vertical cutoff, $\gamma=90\deg$ at altitude of 
40 km.

\begin{figure}[htb]
\plotone{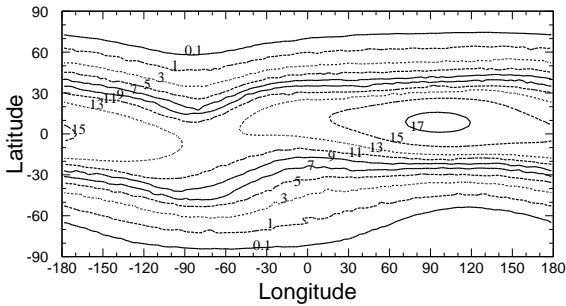}
\label{fig:rigcut}
\caption{The rigidity cutoff in GV/c of normal incident at altitude of 40km.}
\end{figure}

\subsection{Motions of trapped particles in geomagnetic field}
For particles trapped inside the geomagnetic field, there are three types of 
motions (Walt 1994). They are the gyrations along a guiding center magnetic 
field line, the north-south bounce motions due to the repelling force of 
convergent magnetic fields, and the east-west drift motions due to gradient 
drift and curvature drift. The positive charged particles drift eastward while
the negative charged particles drift westward. 

\section{Search for antihelium}
\subsection{Resolution}
After the AMS01 flight, the detector was re-examined, first using the 
heavy ion (He,C) beam from 1.0 to 5.6 GV in GSI-Darmstadt, and then using 
the proton and pion beam 2 to 14 GV in CERN. The performance of the detector 
remained the same before, during, and after the flight (Alcaraz 1999).

At 2 to 10GV, the resolution is approximately 0.11- 0.14. At lower rigidity, 
the resolution becomes worse due to scattering, while at higher rigidity, the 
resolution deteriorates due to limited resolution of silicon tracker.

\subsection{Data selection}
Major contaminations of antihelium are confusions of charge magnitude $|z|$ and
 charge sign. The probability of confusion of $|z|=2$ from $|z|=1$ was 
estimated to be less than $10^-7$ by the measurements of $z^2$ from the TOF and
 tracker. Particle directions are determined by the TOF. 
The large-angle nuclear scattering events were identified and excluded by 
the large error or asymmetry of rigidity. The asymmetry $A_{12}$ is the 
relative error of rigidity of first half track $R_1$ and last half track 
$R_2$, where $A_{12}=(R_1-R_2)/(R_1+R_2)$. 
In addition, events with collinear delta rays were rejected by identifying 
excess of energy within 5mm of the track. Finally, a probabilistic function 
was constructed from measurements of velocity, rigidity, and energy loss which 
described the compatibility of these measurements with passage of helium or 
antihelium. 

\subsection{Antimatter search result}
The last four candidates fail to be compatible with antihelium, and 
$2.86\times 10^6$ helium with rigidity of 1 to 140GV survives all the cuts. 
The antimatter limits at 95\% confidence level are then estimated by the 
following three methods. (1) Assuming that antihelium has the same spectrum as 
helium, an assumption commonly used in many antimatter experiments, the 
$\overline{\mbox{He}}/{\mbox{He}} = 1.1\times 10^{-6}$ 
in the rigidity range of 1 to 140 GV. This result and some previous limits are 
plotted in Figure 4. 
(2) Assuming uniform $\overline{\mbox{He}}$ rigidity spectrum, the 
$\overline{\mbox{He}}/{\mbox{He}}=1.8\times 10^{-6}$ at 1.6 to 40GV and 
$3.9\times 10^{-6}$ at 1.6 to 100 GV. 
(3) For a conservative upper limit, which is independent of the
${\overline{\mbox{He}}}$ spectrum, the $\overline{\mbox{He}}$ limit is a 
function of rigidity (Fig. 9 of Alcaraz 1999). 
  
\begin{figure}[htb]
\plotone{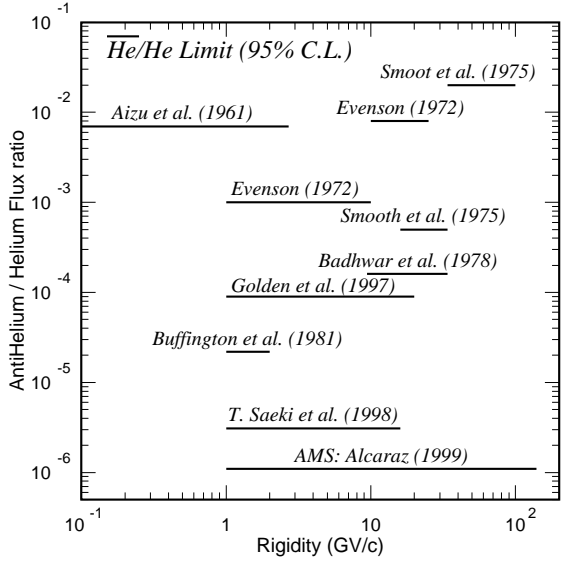} 
\label{fig:all_ahe}
\caption{The AMS antihelium limit is plotted with some previous measurements. 
This limit assumes that antihelium has the same spectrum as helium.}
\end{figure}

\section{Proton spectrum}
Although the primary cosmic ray flux has been measured many times, the AMS 
is the first instrument that measures cosmic rays globally. This information 
is essential to the atmospheric neutrino calculation. 

\subsection{Proton spectrum}
In this study (Alcaraz 2000A), the acceptance was restricted to events with 
incident angle within $32\deg$ from the $\widehat{z}$ axis, the cylindrical 
axis of the AMS01 magnet. Two periods of data were used. For the first 
period, the $\widehat{z}$ axis points to $1\deg$ around the zenith. Data taken 
in this period are referred to as {\em "downward"} going. For the second period,
 the $\widehat{z}$ axis points to $1\deg$ around the nadir. Data taken in this 
period are referred to as {\em "upward"} going. 

The major contaminations of proton are charged pions and deuterons. The pions 
are produced in the top part of the AMS. They account for 5\% at below 0.5 GeV 
and decrease rapidly at higher energy. The deuterons abundance is about 2\% in 
cosmic rays. Both pions and deuterons can be rejected by requiring the measured
 mass to be consistent with mass of proton within three sigmas. Owing to the 
detector resolution and possible energy loss in detector materials, the 
measured spectrum need to be unfolded to get the incident spectrum. This 
procedure uses detector resolution from Monte-Carlo simulation and Bayes 
theorem (reference 7,8 of Alcaraz 2000A). Figure 5 shows the 
differential spectrum of several magnetic latitude intervals.

\begin{figure}[htbp]
\plotone{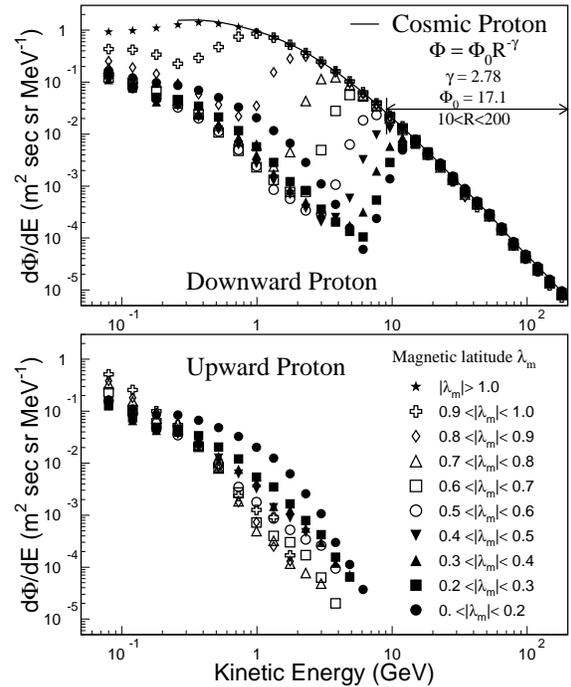}
\label{fig:pflux}
\caption{The upper figure shows the downward proton fluxes in 10 
latitude intervals. The solid line is the cosmic proton. For each latitude 
interval, the proton fluxes have a dip due to the rigidity cutoff; below this 
cutoff, there is a second spectrum. The lower figure shows the upward proton 
fluxes. Cosmic rays do not exist in these upward events.}
\end{figure}

For downward events, the outer envelop is the cosmic ray flux. The proton 
fluxes decrease rapidly in some regions because of the geomagnetic rigidity 
cutoff. However, the proton fluxes rise again at energy lower than the cutoff. 
For upward events, all the protons are below the cutoff. Those spectra below 
cutoff are referred to as {\em "second spectra"}. Since the radiation belts are 
at altitude over 1000km, such large second fluxes are not expected at the AMS 
altitude of around 380km. These second spectra are discussed in detail in 
Section 6.

\subsection{Cosmic proton spectrum}
For primary cosmic ray proton, all the available data are used in a separate 
study (Alcaraz 2000C). The data were collected in three periods in which the 
AMS $\widehat{z}$ axis pointed to $0\deg,\ 20\deg,\ 45\deg$ around the zenith. 
The acceptance of AMS was extended to $38\deg$ from the $\widehat{z}$ axis of 
AMS. The cosmic proton must have rigidity R, such that 
$R > R_c \times (1.2 + 2 \sigma (R_c) )$ where $R_c$ is the maximum 
rigidity cutoff of all incident angles and $\sigma(R_c)$ is the uncertainty of 
rigidity at $R_c$. The acceptance of AMS is $ 0.15 \mbox{m}^2 \mbox{sr}$ on 
average and only weakly dependent on momentum. The background rejection and 
spectrum unfolding are the same as described in the previous section. 

	The systematic errors were studied in detail in this study. The first 
source of systematic error was the variation in trigger efficiency and 
reconstruction accuracy. The total error from this origin is 3.5\%. 
The second source of systematic error came from Monte Carlo corrections. 
This source contributes 3\% in total. The third source of systematic error came
 from the unfolding procedures, they are typically 1\% below 20GeV 
and reach 5\% at 100GeV. The final spectrum is fitted to the power law 
spectrum at rigidity $10<R<100$ GV.
\begin{equation}	d\phi/dR = \phi_0 \times  R^{-\gamma}	\end{equation}
The differential spectrum index $\gamma$ is 
$2.78 \pm 0.009 \mbox{(fit)} \pm 0.019 \mbox{(sys)}$ 
and the normalization constant $\phi_0$ is 
$17.1 \pm 0.15 \mbox{(fit)} \pm 1.3 \mbox{(sys)} \pm 1.5 (\gamma)$ 
 $\mbox{GV}^{2.78} /(\mbox{m}^2 \mbox{sr} \mbox{MeV})$. 
Figure 6 shows the above result and several recent measurements 
and spectrum used in the atmospheric neutrino calculation model. The AMS 
spectrum is consistent with that of previous measurements; however, the HKKM 
model (Honda 1995) seems to have higher flux at energy above 20 GeV.

\begin{figure}[htbp]
\plotone{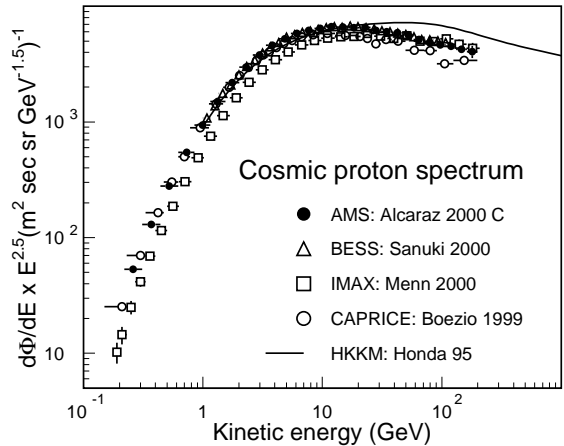}
\label{fig:allp}
\caption{The cosmic proton fluxes measured by the AMS are plotted with some 
previous measurements. The solid line is the primary proton flux used in the 
HKKM atmospheric neutrino model.}
\end{figure}

\section{Leptons}
\subsection{Events selection and contamination}
To study electrons and positrons (Alcaraz 2000B), the track must pass 
through the aerogel Chrenkov counter; therefore the acceptance was limited to 
$25\deg$ from the $\widehat{z}$ axis. All four periods of shuttle attitude, 
$0\deg,\ 20\deg,\ 45\deg$, and $180\deg$ from the 
zenith, are included. The electron candidates are selected with charge -1, 
velocity compatible with speed of light. The major backgrounds are protons with
 wrongly measured rigidity and pions produced in detector materials. These two 
backgrounds are removed by $\chi^2$ of trajectory fitting and number of hits 
near reconstructed track. After this cut, the chance of a proton being 
misidentified as an electron is in the order of $10^{-4}$. The effective energy
 range for electron is 0.2 to 40GeV. The positron candidates are selected with
 charge +1, velocity compatible with speed of light. The major background is 
proton with poorly reconstructed velocity. Above 1 GeV, the protons are 
rejected by requiring the two measurements at two separate Chrenkov counter 
layers to be compatible with those of positrons. Lower energy protons are 
rejected by requiring velocity measurement in TOF and tracker to be compatible 
with that of positrons. Owing to the under-performance of aerogel counter, the 
effective energy range for positron is only 0.2 to 3 GeV.

\subsection{Spectrum}
The conversion from number of events to spectrum is similar to the process used
 in obtaining the proton spectrum. Figure 7 shows the electron 
and positron spectra at several magnetic latitude intervals. Similar to the
proton spectrum, primary cosmic rays exist at high energy and the second 
spectrum appears below the rigidity cutoff.
\begin{figure}[htbp]
\plotone{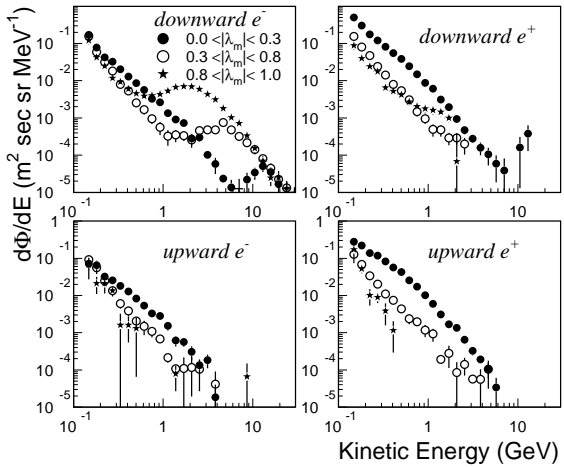}
\label{fig:epflux}
\caption{The electron and positron fluxes measured by the AMS are plotted in 
three latitude intervals. }
\end{figure}

Figure 8 shows the cosmic positron fraction, 
$e^+/(e^+ + e^-)$. The results are consistent with most previous measurements. 
With the current detector, the AMS01 cannot identify the possible positron 
signal from annihilation of WIMP at higher energy. The new AMS detector 
for 2003 will add ring imaging Chrenkov detector and calorimeter and should 
have better chance to detect this dark matter signal.

\begin{figure}[htbp]
\plotone{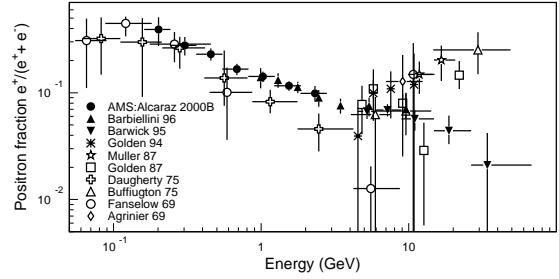}
\label{fig:e+ratio}
\caption{The positron fraction of primary cosmic rays measured by the AMS and 
some previous measurements. At energy $<3$ GeV, the AMS and CAPRICE 
(Barbiellini 1996) show consistent results.}
\end{figure}

\section{Albedo Particles \label{sec:albedo}}
This section contains the preliminary results from the study of the particles 
below the cutoff by the author. The discussion is purely the author's opinion, 
{\bf NOT} that of the AMS collaboration.

\subsection{Introduction}
Below the geomagnetic cutoff, charged particles cannot escape from the 
confinement of the geomagnetic field. Trapped particles in radiation belts have
 the lowest altitude above the atmosphere; they could exist for a long time, 
which is much longer than the drift period. Particles with the lowest altitude 
inside the atmosphere will be absorbed in a short time. Particles below the 
rigidity cutoff are observed in all areas covered by the AMS 
and all the particles studied so far include proton, electron, positron and 
helium. These particles, known as albedo particles, are found to originate from
the atmosphere and rebound to space. The albedo particles had been previously 
detected by many balloon experiments (Bleeker 1965). The splash albedo and 
re-entrant albedo were part of background of cosmic ray antiparticle 
measurements. Most of the radiation belts experiments in the 60s and 70s could
not distinguish between electrons from positrons. However, the presence of 
positrons in the radiation belts had been reported as early as 1983 
(Just 1983, Galper 1983).  

The balloon experiments are mostly conducted in high latitudes and their 
operations were short, from several hours to several days. Although the 
space instruments have longer operation time, their detectors are much smaller 
than the balloon instruments, therefore with smaller acceptance. The AMS 
combines the advantage of these two types of experiments, a large acceptance 
and long duration flight.  The AMS also performs measurements at lower latitude
 and covers 78\% 
of the Earth surface. These factors make AMS a better tool for studing albedo 
particles in detail. AMS also make measurements at higher energy ($\sim$ GeV), 
almost one order of magnitude higher than that in previous radiation belts 
experiments.

Figure 9 shows the AMS results and some previous 
measurements. Only the AMS measurements are shown with error bars, while others
 are shown without. The albedo electron fluxes are quite consistent with 
those of other results. However, variations among the different measurements 
are around one order of magnitude. The reason behind this large difference is 
still unknown; it may be due to the effect of the geomagnetic field, change in 
primary cosmic rays fluxes, or some other systematic effects.

\begin{figure}[htbp]
\plotone{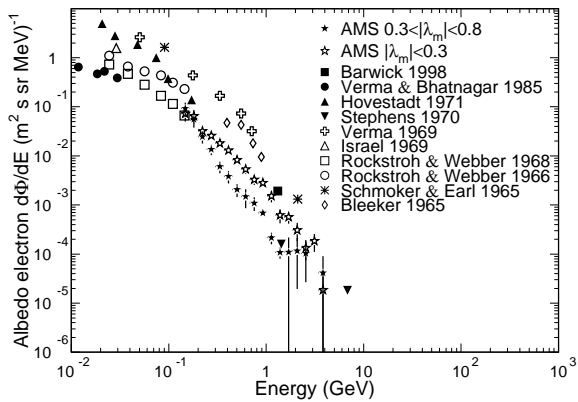}
\label{fig:alb_e}
\caption{The AMS upward going electrons consist only of albedo particles 
(Alcaraz 2000B). The flux of low latitude, $|\lambda_m|<0.3$, and mid 
latitude, $0.3<|\lambda_m|<0.8$ are shown here with some previous 
measurements.}
\end{figure}

The positron and anti-proton can be produced in the atmosphere and rebound to 
space. These particles could be confused with the cosmic rays. Their 
trajectories can be used to distinguish their source. We use a trajectory 
tracing program (Huang 2001) to calculate their past and future trajectories. 
This program uses the Runge-Kuta 4-th order integrator to solve the Lorentz 
equation. The particles are traced backward in time until they exit the 
geomagnetic field, (altitude $>10 R_E$), hit the ground, or exceed a 
pre-determined time limit (typically 10 seconds). Particles coming from 
outside the geomagnetic field are cosmic rays, and have rigidity above the 
cutoff value. Those particles which hit the ground when traced backward are 
referred to as albedo particles. The positions where they come from (hit 
40 km altitude or pass $10R_E$) are called the source. The positions 
where they stop (hit 40 km altitude or pass $10 R_E$) are called the 
sink. The time from source to sink is defined as the flight time. 

Events can be separated into five types according to their source and sink. 
The cosmic rays come from space and sink to the atmosphere. The albedo 
particles come from the atmosphere. Some cosmic rays enter the AMS altitude, 
and are then go back to space; they are cosmic rays but reflected back to 
space by the geomagnetic field. Some albedo events have rigidity above the 
cutoff and could exceed the $10R_E$ limit in forward tracing. These escaped 
albedo events are found in high latitude region where the cutoff values are 
low. Figure 10 shows the trajectory of 
these four types of events. For the trapped radiations, the particle must 
remain inside altitude 100km to $10R_E$ for over 20 seconds. 
\begin{figure}[ht]
\plotone{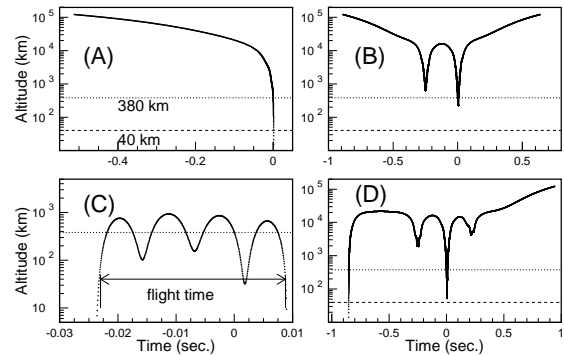}          
\label{fig:traj4}
\caption{The time profile of radial distance of four types of events. 
(A) cosmic rays, (B) reflected cosmic rays, (C) trapped albedo, and 
(D) escaped albedo. Negative time means that the particle is traced backward 
in time, while positive time means that the particle is traced forward in time.
 The 380 km is the mean altitude at the detection site. The 40 km is used to 
define the flight time as shown in (C)}
\end{figure}

We traced all the electrons/positrons and protons below 6 GeV. They all 
have similar distributions. Most of these particles are either cosmic rays of 
trapped albedo particles, few are reflected cosmic rays or escaped albedo. No 
trapped radiation is observed within the AMS altitude and acceptance. However, 
trapped radiation is not totally ruled out, because the AMS does not cover 
the near horizon events.

\subsection{Origin of albedo particles}
At 1 MeV, the neutron could travel $37 R_E$ during its lifetime. The decay 
products, protons and electrons, have few chances to exist inside $10R_E$. The 
cosmic ray albedo neutron decay, CRAND, could not account for the high energy 
albedo particles seen by the AMS. The source of albedo particles must be 
distributed globally and capable of producing positrons in large amount and 
continuously. The most possible mechanism behind the global production of 
positrons is the interaction of primary cosmic rays with atmospheric nucleus. 
\[ \mbox{H, He, ...}\; +\; \mbox{N}_2\;,\; \mbox{O}_2\;,\; ... 
  \rightarrow\; \mbox{H}\; /\; \pi^{0,\pm}\;/\;K^{0,\pm}\; +\; ... \]
\[\pi^{0,\pm}\;,\;K^{0,\pm}  \rightarrow  e^+\;,\; e^- + ... \]
Under some special conditions, these secondary protons, electrons, and 
positrons move upward and become the albedo particles, some could move to the 
AMS altitude and be detected.

\subsection{Flight time}
	Since these albedo particles are below the cutoff, their motions follow
a pattern similar to that of trapped radiations. The bouncing times $N_b$ and 
drift times $N_d$ can be defined by 
\begin{equation}
\begin{array}{ll}
N_b &= \mbox{T} / \tau_b  \\
\tau_b &=0.117 [ 1-0.4635 (\sin \alpha_{eq})^{3/4}] (L/\beta) \\
N_d &= \mbox{T} / \tau_d  \\
\tau_d &=C_d  [1-0.333 (\sin \alpha_{eq})^{0.62}]/(L \gamma \beta^2) \\
    C_d &= 1.557\times 10^4\; \mbox{ for electrons, positrons} \\
       &= 8.841 \; \mbox{ for protons}
\end{array}
\end{equation}
where T is flight time, $L$ is the L-shell number, $\gamma$ is realistic factor
, $\beta$ is velocity, and $\alpha_{eq}$ is the pitch angle at the magnetic 
equator. Note that $\tau_b$ and $\tau_d$ are approximation forms in dipole 
field and accurate to 0.5\% (Walt 94). When the particle rigidity is close to 
the cutoff, the trajectory becomes irregular and these approximation formulas 
could have larger error. 

Figure 11 shows the $N_b$ and $N_d$ distributions of electrons. 
The two horizontal bands are the events that have flight-time $\sim 1/2\tau_b$ 
or $ \tau_b$, referred to as short flight-time (SFT) particles. The two 
vertical bands are the events that have flight-time $\la \tau_d$, referred to 
as long flight-time (LFT) particles. 
\begin{figure}[htbp]
\plotone{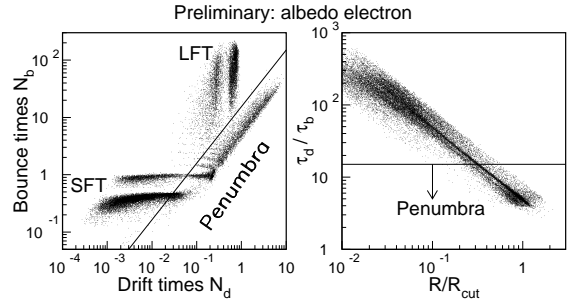}
\label{fig:nbvsnd}
\caption{The left figure shows the distribution of bouncing times and drift 
times. The right figure shows the relation of the $\tau_d/\tau_b$  and ratio of
 rigidity ($R$) and cutoff ($R_{cut}$). When the rigidity inside the penumbra 
region $R/R_{cut}\sim 1$, the trajectory becomes irregular. The cut 
$\tau_d/\tau_b>15$ is shown by the solid line.}
\end{figure}
The diagonal band shows the events with 
$\tau_d/\tau_b<15$, referred to as penumbra events, with rigidity inside the 
penumbra. Because of the momentum error, their origin cannot be determined 
precisely. Section 6.7 will discuss the penumbra events in detail. Albedo 
events are selected with $\tau_d/\tau_b>15$ to avoid contamination from 
primary cosmic rays. Figure 12 shows the $N_b$ and $N_d$ 
distributions of albedo events.

\begin{figure}[tbp]
\plotone{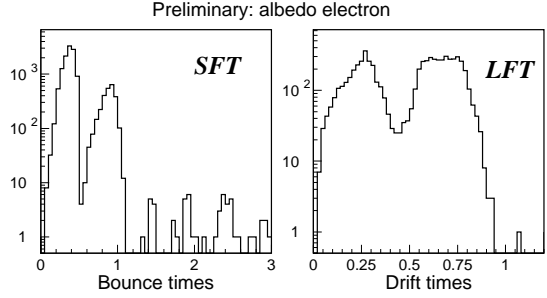}
\label{fig:nbnd}
\caption{The left figure shows the distribution of bouncing times of SFT 
albedo positrons. The right figure shows the distribution of drifting 
times of LFT albedo positrons. Albedo electrons and protons have similar 
distributions.}
\end{figure}

In Fig. 6 of (Alcaraz 2000B), the flight time of LFT particles seems to be 
inversely proportional to kinetic energy. This is because $\tau_d$ is 
proportional to $1/\gamma \beta \simeq 1/\gamma$. 
The flight time of SFT particles seems to be independent of kinetic energy and 
forms two horizontal bands.  The reason is that $\tau_b$ is proportional to 
$1/\beta$. For e+,e- at low latitude, the pitch angel at the magnetic equator 
does not change too much and $\beta \simeq 1$, therefore the flight time 
remains constant for different kinetic energy. If events from all latitudes 
are included, the two bands will be smeared and connected (Fig. 7 of 
Huangmh 2000). For the albedo protons, $\beta$ changes with kinetic energy. 
Therefore, the flight time is not the best indicator for classifying the two 
groups; instead, $N_b$ can distinguish clearly SFT and LFT particles of 
different mass and latitude. In this study, $N_b<3$ indicates SFT 
particles, and $N_b\ge3$ indicates LFT particles.

 Figure 13 shows the flight time distribution of electrons and 
positrons.  The albedo events clearly show a two-peak structure, which 
correspond to SFT and LFT, respectively. The separation point is around 0.3 
second. The albedo proton has a similar two-peak distribution but different 
separation time. 
\begin{figure}[htbp]
\plotone{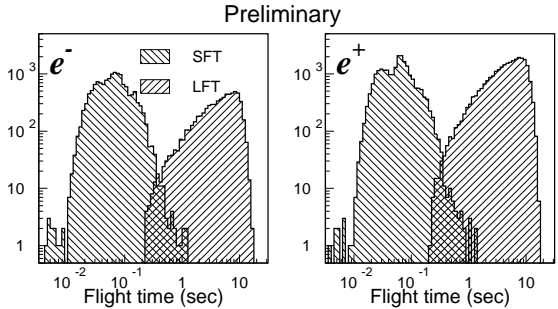}
\label{fig:epft}
\caption{The flight time distributions of albedo electrons (left figure) and 
positrons (right figure), which have $\tau_d /\tau_b>15.$.}
\end{figure}

\subsection{Comparison with radiation belts}
The major differences between the albedo particles and trapped radiations are
 mirroring altitude and life time outside the atmosphere. The mirroring 
altitudes of albedo particles are inside the atmosphere. Therefore, the life 
time is in the order of bounce period for SFT or drift period for LFT. The 
mirroring altitudes of trapped radiations are well above the atmosphere. 
Therefore, the trapped radiations can survive for a long time $\gg \tau_d$. 

Despite these two differences, the albedo particles and trapped radiations 
share some common features, such as trajectory shape and spatial coverage.
Figure 14 shows the distribution of L-shell of SFT and LFT 
electrons. The SFT particles covers L-shell up to L=6.4, where the outer 
radiation belt is. Most of the LFT particles cover L-shell in L=2, where the 
inner radiation belt is. It is clearly seen that these high energy albedo 
particles cover similar space as trapped radiations. 
{\em The radiation belts consist of not only trapped particles 
but also high energy secondary albedo particles.} 
Since these albedo particles not only gyrate 
around the AMS altitude, but also move to higher altitude. 
{\em It is misleading to call these particles ``a ring of 
particle around 380km'' or the third radiation belt.} 

\begin{figure}[htbp]
\plotone{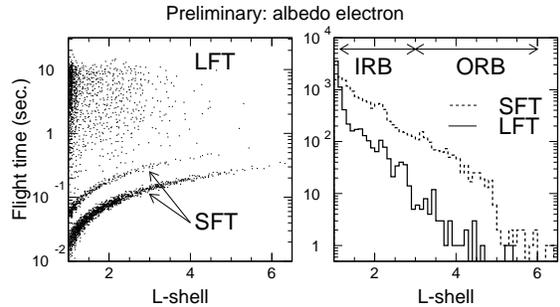}            
\label{fig:lshmom}
\caption{The left figure shows the distribution of flight time as a function of
 L-shell. The right figure shows spatial coverage of albedo particles. The LFT 
particles cover mainly in the inner radiation belt (IRB) and SFT particles 
extend to outter radiation belt (ORB).}
\end{figure}

\subsection{Source distributions}
\subsubsection{Short flight time (SFT)}
The SFT particles are related to the bouncing motions between the 
northern and southern hemispheres, typically crossing the magnetic equator 
once ($N_b\la 1/2$) or twice ($1/2<N_b\la1$) (Huang 2000). 
Figure 15 shows two examples.
The source and sink of SFT particles are near the magnetic footprints of the 
magnetic field lines at the detection sites; therefore they are distributed 
uniformly and show a similar pattern as the flight path of the shuttle. 
Figure 16 shows the source of SFT electrons and positrons.

\begin{figure}[htbp]
\plotone{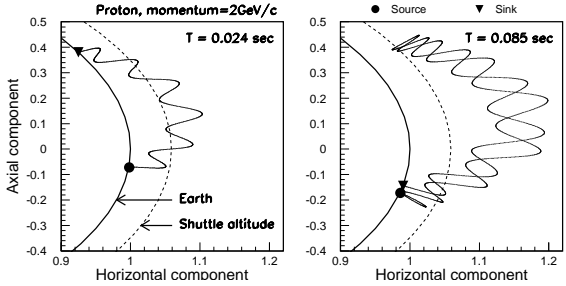}            
\label{fig:sft_trj}
\caption{The typical trajectories of SFT particles are bouncing
between the northern and southern hemisphere, their fight time are in the 
order of 1/2 (left figure) and 1 (right figure) bounce period.}
\end{figure}

\begin{figure}[htbp]
\plotone{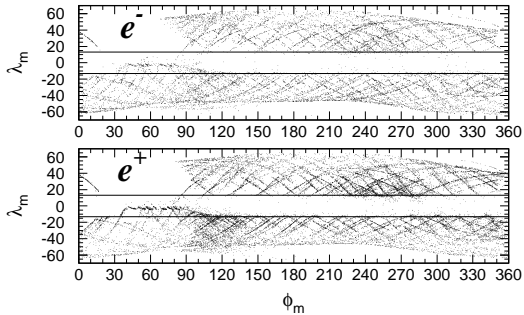}            
\label{fig:sft_src}
\caption{The source position of SFT particles in magnetic coordinates, where  
$\lambda_m$ is the magnetic latitude and $\phi_m$ is the magnetic longitude. 
The two lines mark the equaotrial gap at $|\lambda|=13\deg$. The voids in the 
upper left corner are caused by the deletion of data taken when the AMS was 
inside the SAA region.}
\end{figure}

\subsubsection{Long flight time (LFT)}
The LFT particles are related to the longitudinal drift. Positive charged 
particles drift westward, origin from the magnetic western hemisphere 
($\phi_m=180\deg$ to $360\deg$) and sink to the magnetic eastern hemisphere
($phi_m=0\deg$ to $180\deg$).  The typical drift times are $N_d\sim$1/4 
and 3/4. Figure 17 shows the simplified trajectory of two 
examples. A few of the LFT events just drift $N_d \sim 1/8$ and do not cross 
$\phi_m=180\deg$, these events are referred to as intermediate flight-time 
(IFT)events.

\begin{figure}[htbp]
\plotone{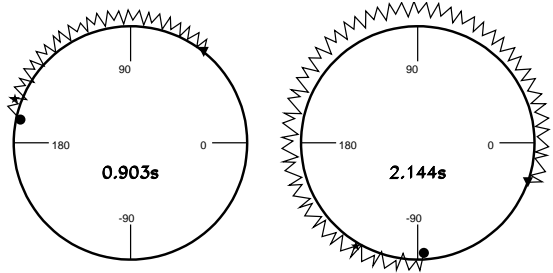}            
\label{fig:lft_trj}
\caption{Two simplified trajectories of LFT particles viewed from North pole. 
The circle is the Earth and the azimuth angles is the geographic longitude. 
Only the mirroring points (which are near the Earth) and points when crossing 
the magnetic equator (which are away from the Earth and have highest altitude 
within $\tau_b/2$) are plotted. The LFT particles drift eastward/westward, 
their typical fight time are in the order of 1/4$\tau_d$ (left figure) and 
3/4$\tau_d$ (right figure). The number shown in the center of the figure is the
 flight time in seconds.} 
\end{figure}

The source (sink) position of these LFT particles concentrate in 
two regions around the SAA. Figure 18 shows the source of LFT 
electrons and positrons. Owing to the charge-time symmetry, the sinks of 
$e^+(e^-)$ are the sources $e^-(e^+)$. Albedo protons and helium 
have similar distribution as positrons.
\begin{figure}[htbp]
\plotone{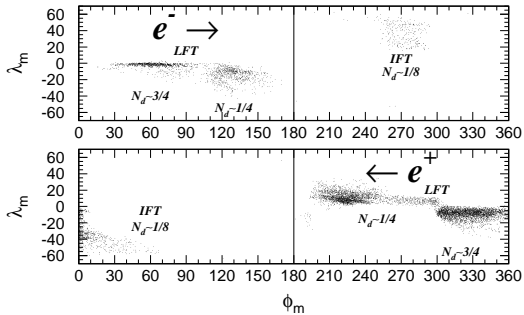}            
\label{fig:lft_src}
\caption{The source position of LFT particles in magentic coordinates. The 
arrows show the drift direction. The typical sources are concentrated in two 
areas near $\phi_m=0\deg$. The IFT events are in $\phi_m<180\deg$ for 
electrons and $\phi_m>180\deg$ for positrons.}
\end{figure}

LFT particles originate near $\phi_m=0\deg$ where the surface magnetic 
fields are the weakest. Drifting away from the SAA, LFT particles encounter 
stronger magnetic field and mirroring at higher altitude. When they pass the 
$\phi_m=180\deg$, the surface equatorial magnetic field are the strongest, the 
mirroring altitude decreases and finally intersects the atmosphere again. The 
source and sink are almost symmetric along $\phi_m=0\deg$, however, the 
geomagnetic field is not just an offset dipole, the multipole moment produce 
two groups of LFT. The offset dipole will produce even distribution in drift 
time distribution, unlike that shown in Figure 12.

The source position of electrons concentrate in two regions in the south 
magnetic hemisphere ($\lambda_m < 0$) and the source of positrons, protons and 
helium are in two regions on both sides of the magnetic equator. This is also 
caused by the multipole moment, which produces asymmetry in the magnetic field 
strength. The particles fall to the side where the magnetic field is weaker.
However, the exact production sites are not only in the 40 km altitude, 
therefore, the exact source positions should be more dispersive.

\subsection{Albedo positron electron ratio}
The flux ratio of positrons to electrons varies with  magnetic latitude and 
can be as large as 4 near the magnetic equator. Some balloon experiments, 
operated in high latitude regions, obtained a ratio of approximately 1. The 
excess of antimatter arouses questions concerning their origin. At high 
latitude, these albedo positrons could be higher than the cutoff and be 
mistaken as cosmic rays. 

Owing to the geomagnetic rigidity cutoff, there are more cosmic rays coming 
from the west than from the east. This east-west effect is 
the major factor behind the large positron electron ratio. Because of the 
geomagnetic field, only the positrons coming from the west and electrons 
coming from the east have the chance to move upward. Since the primary cosmic 
ray fluxes from the west are larger than those from the east, the secondary 
positrons from the west are more abundant than the electrons from the east. 
The difference in rigidity cutoff decreases with increasing magnetic latitude, 
so does the flux ratio.

A simple model was proposed to explain this large $e^+ / e^-$ 
ratio, shown in Figure 19 (Huang 2000). Although this is just a first-order approximation, 
the quantitative agreements evidence the validity of this model.

\begin{figure}[htbp]
\plotone{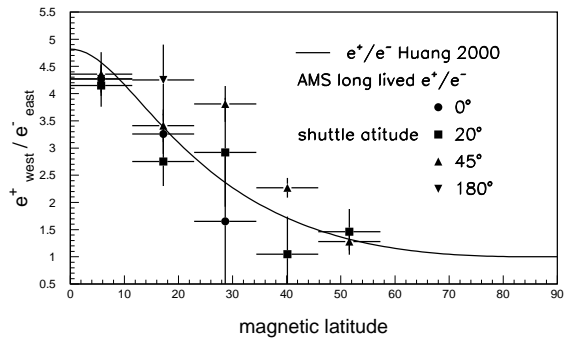}            
\label{fig:rep}
\caption{The AMS long flight-time $e^+ / e^-$ can be explained by east-west effect.}
\end{figure}

\subsection{Penumbra events}
Events with $\tau_d/\tau_b<15$ originate mostly in high latitude region or 
have high rigidity. Figure 20 shows some plots of these 
events. Their flight-time distributions do not have two distinct peaks, 
although the $N_b$ can still be used for classifying SFT and LFT. Some LFT 
particles can drift over several drift periods. This is the typical behaviors 
of the L-shell splitting at L-shell $>$4. Also, the rigidity of these events 
are inside the penumbra, their trajectories become irregular. The source of 
SFT particles is also uniformly distributed. However, the void in 
Fig. \ref{fig:sft_src} does not appear in this figure. Unlike those in the 
Fig. \ref{fig:lft_src}, the sources of LFT particles spread over all 
longitudes. These penumbra events have different distributions. 
{\em The LFT penumbra events are not included in } (Alcaraz 2000B).

\begin{figure}[ht]
\plotone{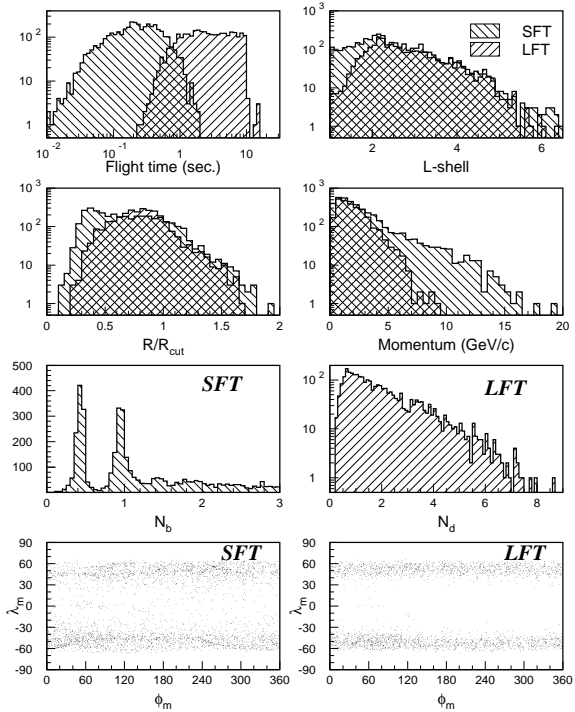}            
\label{fig:highlat}
\caption{The various distributions of events with  $\tau_d/\tau_b<15$. 
These events originate mostly and are detected in high latitude and have 
rigidity close to the cutoff. Many of these particles can drift over one 
circle. Their source distributions are different from those of
 events with $\tau_d/\tau_b>15$.}
\end{figure}

\section{Summary}
The physics results of the AMS01 shuttle flight are summarized as follows. 
\begin{itemize}
\item Antimatter limit: $1.1\times 10^{-6}$ for rigidity 1. to 140GV/c$^2$.
\item Primary cosmic ray spectrum in 10GeV $<$ R $<$ 200GeV, 
	$\Phi(R)=\Phi_0 R^{-\gamma}$ where \\
	$\Phi_0=17.1\pm2.0$GeV$^{2.78}$/(m$^2$ s sr MeV) \\
	and $\gamma=2.78\pm0.021$.
\item Measurements of primary electrons and positrons are consistent with 
	previous results.
\item Many particles are discovered below the geomagnetic rigidity cutoff, 
	these particle must originate in the atmosphere and then rebound to 
	space, i.e. atmospheric albedo particles.
\item For particle well below the cutoff, two groups of atmospheric albedo 
	particles are found, short flight-time and long flight-time particles.
	\begin{itemize}
	\item Short flight-time particles are related to bounce motions,  
		their sources are uniformly distributed, and they cover L-shell
		up to 6.
	\item Long flight-time particles are related to longitudinal drift 
	motions, their source/sink positions are distributed at two 
	distinct areas at east/west of the magnetic prime-meridian 
	($\phi_m=0\deg$), and they cover L-shell up to 4.
	\item A large latitude-dependent albedo e+/e- ratio is discovered. 
	A possible explanation is the east-west effect.
	\end{itemize}
\item Particles near the cutoff exist mostly in high latitude and could 
	drift over several drift periods. Their source distributions are 
	different from those of particles well below cutoff.
\end{itemize}

\acknowledgements
The author wishes to thank the organizer of this workshop, Professor C.M. Ko 
for his kind invitation. This study was supported by grant 
NSC89-2811-M-001-0076 from the National Science Council, Taiwan, ROC.

\end{document}